\documentclass[preprint,aps]{revtex4}
\usepackage{bm}
\usepackage[dvips]{graphicx,color}
\usepackage[latin1]{inputenc}
\begin{document}
\title{Monte-Carlo simulation for fragment mass and kinetic energy distributions from neutron induced fission of  $^{235}U$}

\author{M. Montoya $^{a,b}$, E. Saettone $^{b}$, J. Rojas $^{a,c}$}
\email{jrojast@unmsm.edu.pe}
\affiliation{$^{a}$ Instituto Peruano de Energ\'{\i}a Nuclear, Av. Canad\'a 1470, Lima 41, Per\'u.}
\affiliation{$^{b}$ Facultad de Ciencias, Universidad Nacional de Ingenier\'{\i}a, Av. Tupac Amaru 210, Apartado 31-139, Lima, Per\'u.}
\affiliation{$^{c}$ Facultad de Ciencias F\'{\i}sicas, Universidad Nacional Mayor de San Marcos, Av. Venezuela s/n, Apartado Postal 14-0149, Lima~-~14, Per\'u.}

\begin{abstract}
The mass  and kinetic energy  distribution of nuclear  fragments from thermal neutron induced  fission  of  $^{235}U$  have been  studied  using  a Monte-Carlo  simulation. Besides reproducing  the pronounced broadening on the standard deviation of the final fragment kinetic energy distribution ($\sigma_{e}(m)$) around the mass number $m$ = 109, our simulation also produces  a second broadening around  $m$ = 125, that is in agreement with  the experimental data obtained by  Belhafaf {\it et al}. These results are consequence of  the characteristics of the  neutron  emission, the variation in the primary fragment mean kinetic energy and the yield  as a function of the  mass. \\


Mediante la simulaci\'on con el m\'etodo Monte-Carlo fue estudiado  la distribuci\'on de masas y energ\'{\i}a cin\'etica de los fragmentos  de la fisi\'on inducida por neutrones t\'ermicos del $^{235}U$. Adem\'as de reproducir el ensanchamiento pronunciado en la desviaci\'on est\'andar de la distribuci\'on de la energ\'{\i}a cin\'etica de los fragmentos finales ($\sigma_{e}(m)$) alrededor del n\'umero m\'asico $m$ = 109, nuestra simulaci\'on tambi\'en produce un segundo ensanchamiento alrededor de $m = 125$, en concordancia con los datos experimentales obtenidos por Belhafaf {\it et al}. Estos resultados son consecuencia de las caracter\'{\i}sticas de la emisi\'on de neutrones, la variaci\'on de la energ\'{\i}a cin\'etica media y el rendimiento de los fragmentos primarios en funci\'on de la masa. \\

\textsl{Descriptores}: Monte-Carlo; fisi\'on inducida por neutrones; $^{235}U$; desviaci\'on est\'andar. \\


\end{abstract}

\maketitle
\section{Introduction}
\label{intro}
Since the discovery of the  neutron-induced fission of uranium by Hahn and
Strassmann in  1938~\cite{hahn}, much effort has been made to understand the 
processes involved in it and to  measure the relevant  fission parameters. 
Nowadays  several  aspects of heavy nuclei fission  seem to be clarified. Meitner
and Frisch  suggested  a theoretical  explanation based on  a nuclear
liquid-drop model~\cite{meit}, and, over the past 30 years the model has provided considerable insight into nuclear structure \cite{pomor}. 
It is known that the  de-excitation by fission of heavy nuclei depends of
the quantum  properties of  the  saddle point  and of the  associated fission
barrier. The  detection   of  fission  isomers has  been interpreted  by
the  secondary well  in the fission barrier~\cite{boste}. The nascent
fragments begin  to be  formed at  the saddle  point, then the system falls
down to  the fission valley (energetically preferred paths to  fission)
and  ends at  the scission  configuration  where fragments interact only
by Coulomb force.  Moreover, at scission, the fragments have acquired a 
pre-scission kinetic energy. Over the fission valley,
the system could be   described  by   collective  variables
(such as deformation,  vibration, rotation,  etc.) and intrinsic variables
(such as quasi-particles excitations).
Nevertheless, the  dynamics  of the fission  processes are not yet completely
understood~\cite{schmid}.  In particular, it is neither known the nature of the  coupling  
between the collective and intrinsic degrees  of freedom during  the descend from the saddle to scission, nor known how it does arise.The physics problem of the description of the fission fragment mass and kinetic energy distributions is very closely related to the topological features in the multi-dimensional potential energy surface ~\cite{moller}.
\begin{figure}
\centering
\includegraphics[angle=270,width=0.8\textwidth]{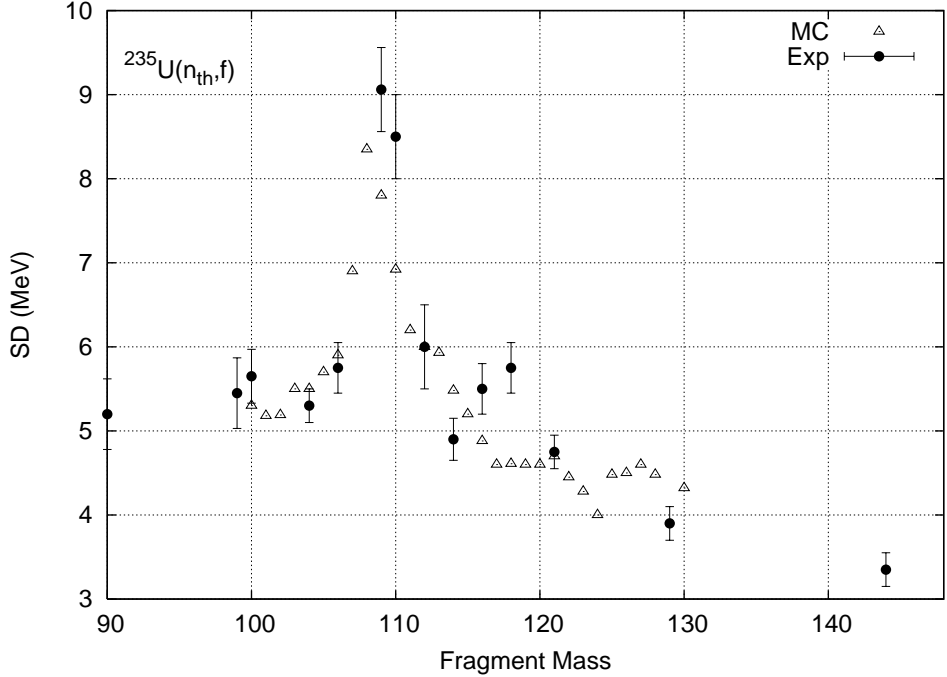}
\caption{Thermal neutron induced fission of $^{235}U$. Standard deviation of the final fragment kinetic energy distribution as a function of the final mass $m$, as a result of Monte-Carlo simulation ($\bigtriangleup$), and  experimental data ($\bullet$). Both from ~\cite{brissot}.}
\label{fig:1}
\end{figure}
In the low-energy fission, several final fragment characteristics can be
explained in  terms  of a  static  scission model  of two  coaxial juxtaposed
deformed  spheroidal fragments, provided  shell  effects, affecting the
deformation energy of the fragments. These shell effects corrections,
determined by the Strutinsky  prescription and  discussed by Dickmann
{\it et al.}~\cite{dick} and Wilkins~\cite{wilkins}, subsequently generate
secondary minima in the total potential energy surface corresponding to fragments having
some particular neutron  or proton shell configurations. If the final
fragment characteristics were  governed by the properties of the fragments
themselves, a basic  argument in any statistical theory, one would  then expect
an  increase  in   the  width  of  the  kinetic  energy distribution
curve for fragment masses $A$, having  the above mentioned special neutron
or proton shell arrangements.
\begin{figure}
\centering
\includegraphics[angle=270,width=0.8\textwidth]{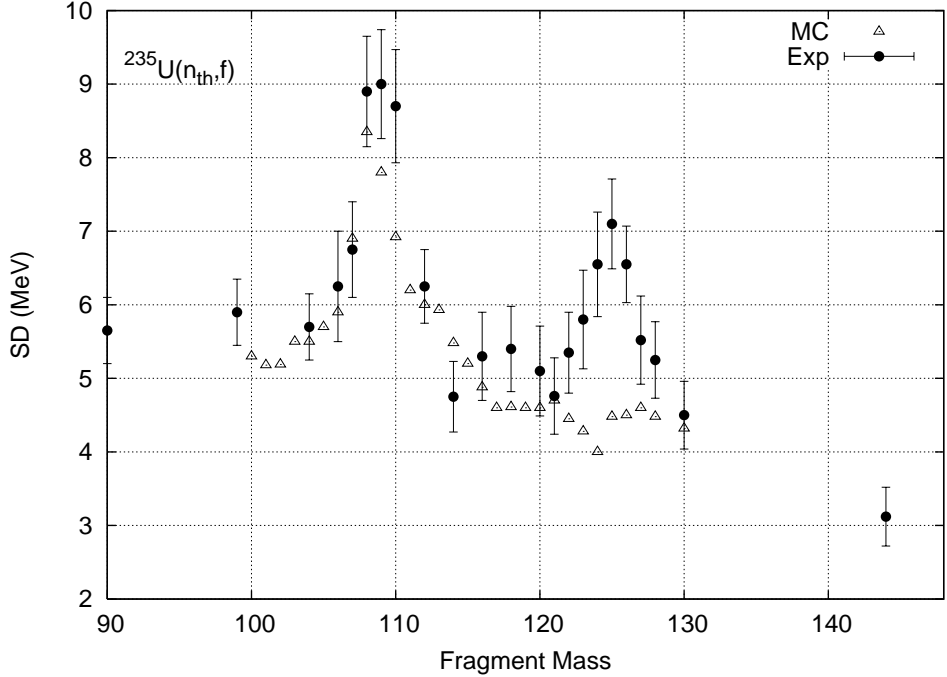}
\caption{Thermal neutron induced fission of $^{235}U$. Simulated standard deviation of the final fragment  kinetic energy distribution as a function of the final mass $m$ ($\triangle$), from Ref. ~\cite{brissot}, does not reproduce the experimental broadening around $m=125$ taken from Ref. ~\cite{belha}.}
\label{fig:2}
\end{figure}
In order to address this question, the  fission parameters of the primary fragments (pre-neutron emission) have been  the most studied are the mass yield  ($Y(A)$) and the kinetic energy ($E(A)$) distribution. Nevertheless, direct measurements can only be  carried out on the final fragments (post neutron emission)  mass yield $Y(m)$ and  kinetic  energy ($e(m)$).
Therefore it  is crucial  to find out what is the relation between  the  primary and the final kinetic energy distributions, as well as the relation between the $Y(A)$ and $Y(m)$ curves.
For  thermal neutron induced fission of   $^{235}U$,  which in fact is the fission of excited $^{236}U$ ($^{236}U^*$) formed by neutron absorption by $^{235}U$, the $e(m)$ distribution was
experimentally determined  by  Brissot { \it et  al.}~\cite{brissot}. This distribution was represented by the mean value of kinetic energy $\overline e$ and the standard deviation  (SD) of  the kinetic energy $\sigma_{e}$ as  function of the final mass  $m$. As seen in Fig. ~\ref{fig:1} the plot of both the measured values and the results of a Monte-Carlo simulation of $\sigma_{e}$ from a primary distribution $E(A)$ without broadenings, shows one pronounced broadening around   $m \approx~109$. This  Monte-Carlo simulation result suggests that the broadening  does not exist on the primary fragment kinetic energy as a function of the primary fragment mass.
In a latter experiment, Belhafaf {\it et al.} ~\cite{belha}, repeated the experiment of Brissot { \it et  al.} for neutron induced fission of $^{235}U$, obtaining a second broadening around $m \approx ~ 125$ (see Fig.2). 
A Monte-Carlo simulation made by these authors, from a  primary distribution of $E(A)$ without a broadening, reproduced the experimental broadening  on $\sigma_{e}$ at $m=109$, but failed to reproduce the broadening around  $m=125$. They suggested that this broadening  must exist in the primary fragment kinetic energy ($E(A)$) distribution,~and accordingly they fitted their experimental data from a distribution with a broadening around  $A$=126.
In this  paper, we present new Monte-Carlo  simulation results for thermal neutron induced fission of $^{235}U$. We compute both the mass and kinetic energy  of the  primary and final  fission fragments, and we show that the broadenings on the  $\sigma_{e}$ curve around  the  final fragment  masses  $m=109$  and $m = 125$ can be reproduced without assuming an adhoc initial structure on $\sigma_{E}(A)$ curve.
\begin{figure}
\centering
\includegraphics[angle=270,width=0.8\textwidth]{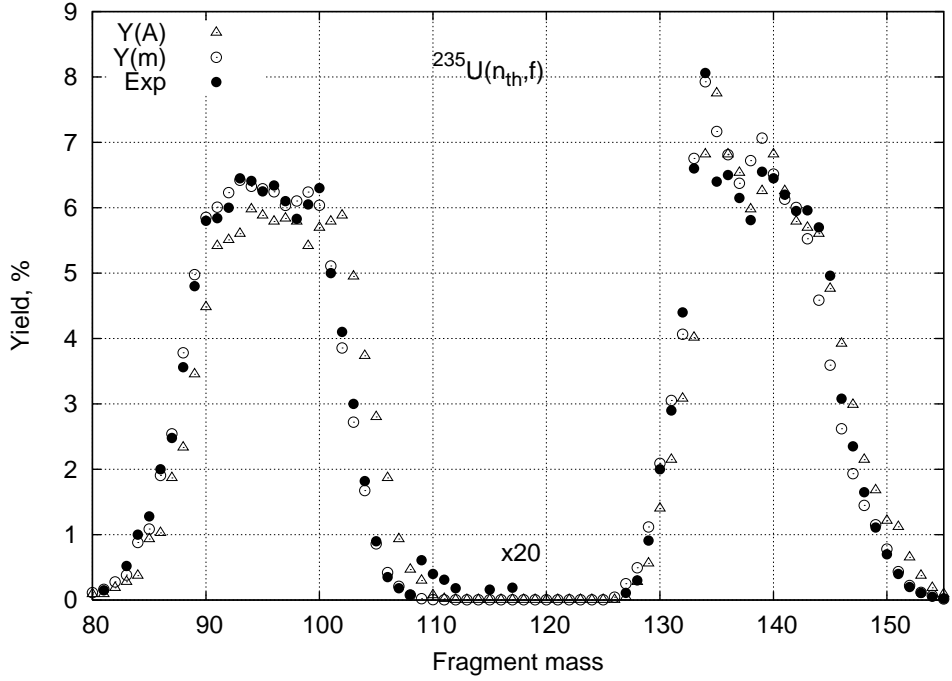}
\caption{Thermal neutron induced fission of $^{235}U$. Simulation results for the primary ($\triangle$) and final ($\odot$) mass yields are presented together with experimental data ($\bullet$), taken from Ref. \cite{Wagemans}}.
\label{fig:yield}
\end{figure}
\section{Monte Carlo simulation model}
\label{sec:model}
\subsection{Fragment kinetic energy and neutron multiplicity}

In the process of thermal neutron induced fission of $^{235}U$, the excited composed nucleus $^{236}U^*$ is formed first. Then, this nucleus splits in two complementary fragments having $A_1$  and $A_2$ as mass numbers, and $E_1$ and $E_2$ as kinetic energies, respectively.

Using relations based on momentum and energy conservation, the total kinetic energy of complementary fragments results
\begin{equation}
TKE = E_1 + E_2 = \frac{A_1+A_2}{A_2}E_1.
\label{eq:TKE}
\end{equation}
The total excitation energy is given by
\begin{equation}
TXE= Q-\epsilon_n - TKE,
\label{eq:TXE}
\end{equation}
where $Q$ is the difference between fissioning nucleus mass and the sum of two complementary fragments masses, and $\epsilon _n$ is the separation neutron energy of $^{236}U$. 
Using equation (\ref{eq:TKE}) in (\ref{eq:TXE}) and taking into account that $A_1 + A_2 = 236$ results
\begin{equation}
TXE = Q + \epsilon_n -\frac{236}{236-A}E,
\label{eq:TXEE}
\end{equation}
where $A$ and $E$ are the mass number and kinetic energy, respectively, of one of the two complementary fragments.
It is reasonable to assume that the excitation energy of one complementary fragment ($E^*$) is proportional to the total excitation energy, then,
\begin{equation}
E^* \propto TXE = Q  + \epsilon_n -\frac{236}{236-A}E,
\label{eq:Ex}
\end{equation}
and that the number ($\nu$) of neutrons emitted by a fragment is proportional to its excitation energy, i.e.
\begin{equation}
 \nu \propto E^*.
\label{eq:nuE}
\end{equation}
From relations (\ref{eq:Ex}) and (\ref{eq:nuE}) one derives a linear relation between $\nu$ and $E$:
\begin{equation}
 \nu = a + bE
\label{eq:nuabE}.
\end{equation}
Taking into account that there is no neutron emission $\nu = 0$ for fragments having the maximal kinetic energy ($E_{max}$)  and assuming that for the average value of fragment kinetic energy $\nu = \bar \nu$, the relation (\ref{eq:nuabE}) results
\begin{equation}
\nu = \bar \nu (\frac{E_{max} - E}{E_{max}- \bar E}).
\label{eq:nuEmax}
\end{equation}
Let be the parameter $\beta$ define the maximal value of kinetic energy by the relation
\begin{equation}
E_{max} = \bar E + \frac{\sigma_E}{\beta}.
\label{eq:Emax} 
\end{equation}
Then, the relation (\ref{eq:nuEmax}) may be expressed as
\begin{equation}
\nu = \bar \nu (1 - \beta (\frac{E- \bar E}{\sigma_E})).
\label{eq:nualphsbetaE}
\end{equation}
Because the neutron number $N$ is integer, it will be defined as the integer part of (\ref{eq:nualphsbetaE}), i.e.
\begin{equation}
N = {\rm Integer ~ part ~ of}(\alpha + \bar \nu (1 - \beta (\frac{E- \bar E}{\sigma_E}))),
\label{eq:nu}
\end{equation}
where $\alpha$ is used to compensate the effect of the change from a real number $\nu$ to an integer number $N$.
\subsection{Simulation process}
In our  Monte Carlo simulation the input quantities are the primary fragment yield ($Y$), the average kinetic energy ($\bar E$), the standard deviation of the kinetic energy distribution ($\sigma_E$) and the average number of emitted neutron ($\bar \nu $) as a function of primary fragment mass ($A$). The output of the simulation for the final fragment are the yield ($Y$), the standard deviation of the kinetic energy distribution ($\sigma_E$) and the average number of emitted neutron ($\bar \nu $) as a function of final fragment mass $m$.

For the first simulation,  we take $Y$ from Ref. ~\cite{Wagemans}, $\bar \nu$ from experimental results by Nishio {\it et al.} ~\cite{Nishio}, and $\bar E$ from Ref. ~\cite{belha}. The first standard deviation $\sigma_E$ curve is taken without any broadening as function of $A$.  Then, we adjust $Y(A)$, $\nu (A)$, $\bar E(A)$ and $\sigma_E(A)$ in order to get $Y(m)$, $\bar \nu $, $\bar e(m)$, $\sigma_e(m)$ in agreement to experimental data.

In the simulation, for each primary mass $A$, the kinetic energy of the fission fragments is chosen randomly from a Gaussian distribution
\begin{equation}
P(E)=\frac{1}{\sqrt{2\pi}\sigma_{E}}
exp\biggl[-\frac{(E-\overline{E})^2}{2\sigma^2_{E}}\biggr],
\label{eq.ETD}
\end{equation}
where $P(E)$ is the probability density of energy with mean value $\overline{E}$  and standard deviation $\sigma_{E}$.

For each $E$ value, the simulated number of neutrons N is calculated with the relation (\ref{eq:nu}). The final mass of the fragment will be, 
\begin{equation}
m=A-N.
\label{eq:mass}
\end{equation}
Furthermore, assuming that the fragments loose energy only by neutron evaporation and not by gamma emission or any other process, and neglecting the recoil effect due to neutron emission, then the kinetic energy $e(m)$ of the final fragment will be given by
\begin{equation}
\label{eq:ef}
e(m)=(1-\frac {N}{A})E.
\end{equation}
With the assemble of values corresponding to $m$, $e$ and $N$, we calculate $Y(m)$, $\bar e(m)$, $\sigma_e(m)$ and $\nu (m)$.

On the other hand, to obtain an acceptable statistics during the simulation, we have considered a total number of fission events of $^{235}U$ of the order of $10^{8}$. At the same time, we have used the Box-Muller method to generate the random numbers with the required normal distribution~\cite{woolf}, and have computed the SD of all the relevant quantities by means of the following expression which for $e(m)$,  read as
\begin{equation}
\sigma^{2}(m)=\frac{\sum_{j=1}^{N_{j}(m)} e^{2}_{j}(m)}{N_{j}(m)}-{\bar e}^2(m),
\label{eq:SD}
\end{equation}
where $\bar e (m)$ is the mean value of the kinetic
energy of final fragments with a given mass $m$, and $N_{j}(m)$ is the number of fission events corresponding to that mass.
\begin{figure}
\centering
\includegraphics[angle=270,width=0.8\textwidth ]{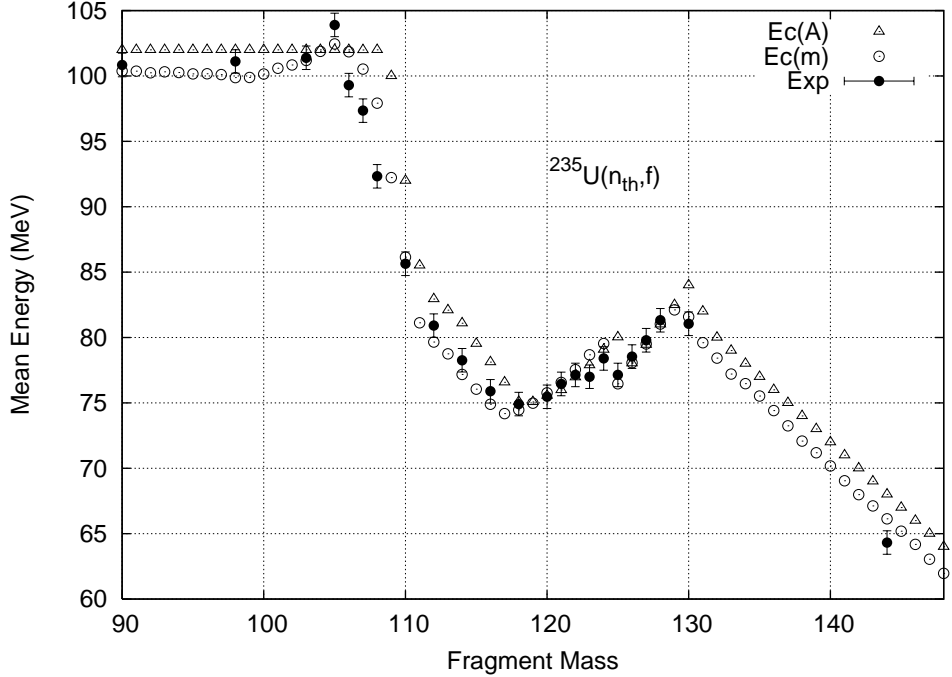}
\caption{Thermal neutron induced fission of $^{235}U$. Mean kinetic energy of the final fragment ($\odot$) and the mean kinetic energy of the primary fragments $\triangle$, as a result of simulation in this work, to be compared to experimental data ($\bullet$) taken from Ref. \cite{belha}.}.
\label{fig:ekm}
\end{figure}
\section{Results and discussion}
\label{results}
The simulated final mass yield curve $ Y(m)$ and the primary mass yield
curve $ Y(A)$ are illustrated  in Fig.~\ref{fig:yield}. As expected, due to neutron emission, the  $ Y(m)$ curve is shifted from $  Y(A)$ towards smaller fragment masses.
\begin{figure}
\centering
\includegraphics[angle=270,width=0.8\textwidth]{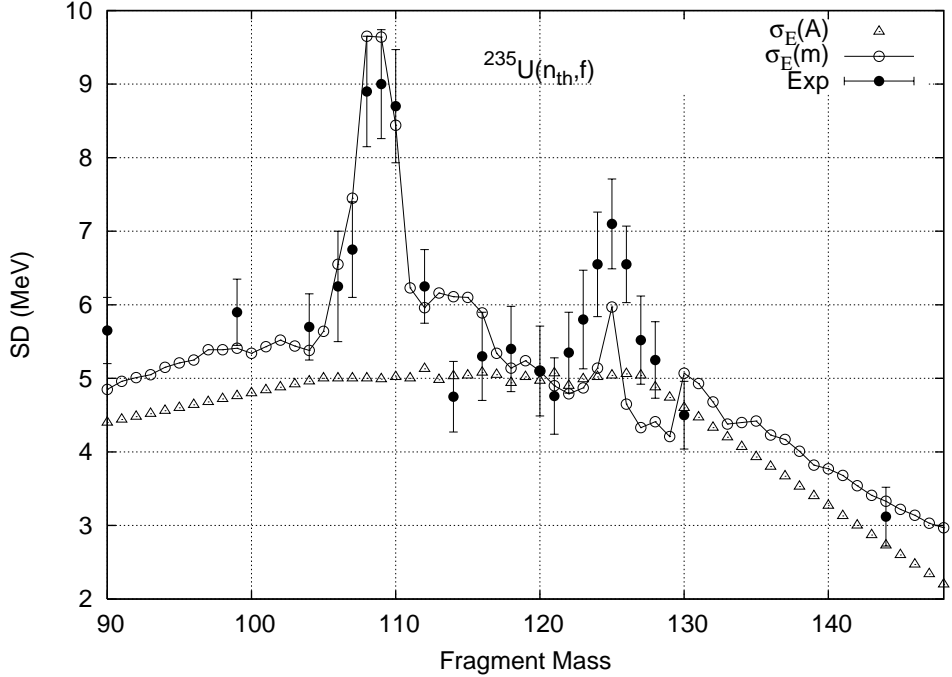}
\caption{Thermal neutron induced fission of $^{235}U$. Standard deviation of final fragment kinetic energy distribution ($\odot$) and standard deviation of primary fragments kinetic energy distribution ($\triangle$), as simulated in this work, to be compared to experimental data ($\bullet$) from Ref. \cite{belha}.}
\label{fig:sde}
\end{figure}
As stated in sect.~\ref{sec:model}, the  primary kinetic energy ($E(A)$) is generated from a  Gaussian distribution, while  the final  kinetic energy ($e(m)$) is calculated  through  Eq.~(\ref{eq:ef}).
The plots of the simulated mean kinetic energy for the primary and final fragments as function of their corresponding  masses, are shown in Fig.~\ref{fig:ekm} . In general, the simulated average final kinetic energy curve as a function of final mass ($\bar e(m)$) have roughly a shift similar to that of $Y(m)$ curve, and a diminishing given by relation (\ref{eq:ef}) with $N=\bar \nu$. The exceptions of this rule are produced in mass regions corresponding to variations of the slope of $Y(A)$ or $\bar E(A)$ curves, for example for $A=109$, $A=125$ and $A=130$.
Furthermore, Fig.~\ref{fig:sde} displays  the  standard  deviation   of  the kinetic energy distribution of the primary fragments and the standard deviation of the kinetic energy of the final fragments ($\sigma_{e}(m)$). The plots of $\sigma_{e}(m)$ reveal the presence  of a  pronounced broadening around  $m$ = 109, and a second broadening is found around $m = 125$, in a mass region where there are variations of the slopes of $Y(A)$ or $\bar E(A)$ curves. There is no experimental data around $m=130$. Nevertheless, if one takes the experimental value $\sigma_e=3.9 MeV$ for $m=129$ from Ref. \cite{brissot} and one puts it on Fig. ~\ref{fig:sde}, the beginning of another broadening for $m=130$ is suggested. 

These results were obtained with a simulated primary fragment kinetic energy distribution (see Fig. \ref{fig:sde},$\triangle$) without broadenings in the range of fragment masses $A$  from 90 to 145. If one simulates an additional source of energy dispersion in $\sigma_E$, without any broadening, no broadening will be observed on $\sigma_e$.

\begin{figure}
\centering
\includegraphics[angle=270,width=0.8\textwidth]{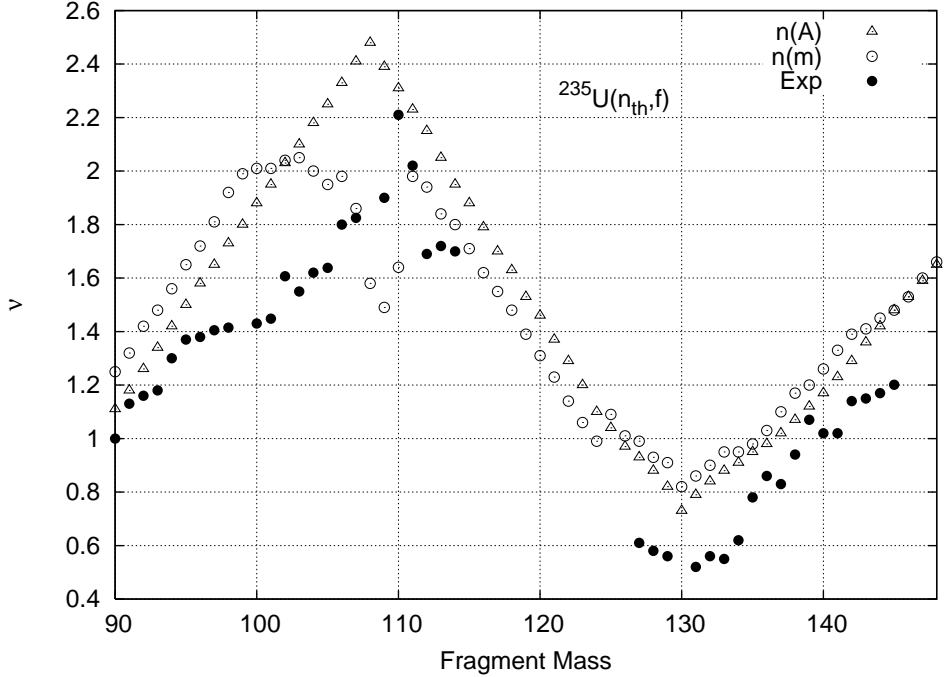}
\caption{The average number of emitted neutrons from fission of $^{235}U$:~ as a function of the primary fragment mass A ($\triangle$), as a function of final fragment mass ($\odot$) both as result of simulation and experimental data ($\bullet$), taken from Ref.~ \cite{Nishio}}
\label{fig:neue}
\end{figure}
Both the shape and height of the broadenings of $\sigma_{e}(m)$ are sensitive to the value of parameter $\alpha$ and $\beta$ appearing in Eq.~(\ref{eq:nu}). A higher value of $\alpha$ will produce a larger broadening of SD. The effect of $\beta$ on broadening depends much on mass region. For the region $m=109$, a higher value of $\beta$ will produce a larger broadening of SD. The simulated results for $\sigma_{e}(m)$ presented in Fig. \ref{fig:sde} were obtained  with $\alpha$ = 0.62 and $\beta$=0.35.
 
The simulated average number of emitted neutron $\bar \nu (m)$ curve is shifted from $\bar \nu (A)$ in a similar way as $Y(m)$ relative to $Y(A)$(see Fig. \ref{fig:neue}).

The presence of broadenings about $m = 109$ could be associated with neutron emission characteristics (approximately $\bar \nu = 2$) and a very sharp fall in kinetic energy from \mbox{$E$ =100 MeV} to $E$ =85.5 MeV, corresponding to $A$=109  and $A$=111, respectively. 
The second broadening is produced by  a discontinuity of the curve $\bar E(A)$ between $A$ =126 to \mbox{$A$ =125}, which is necessary to reproduce a similar discontinuity between $m$ =125 to $m$ =124. We give emphasis to the shape of $\sigma_e$ which increase from $m=121$ to $m=125$ and it decreases from $m=125$ to $m=129$ as occurs with experimental data.
\section {Conclusion}
Using  a simple model for the neutron emission by fragments, we have carried out a Monte-Carlo simulation for the mass and kinetic energy distributions of final fragments from thermal neutron induced fission of $^{235}U$. In comparison with the primary fragments, the final fission fragments have eroded kinetic energy and mass values, as much as to give rise to the appearance of broadenings in the standard deviation of the final fragments kinetic energy as a function of mass $\sigma_{e}(m)$ around $m$ = 109 and $m$ = 125 respectively. These broadenings are consequence of neutron emission and variations on slopes of primary fragments yield ($Y(A)$) and mean kinetic energy $\bar E(A)$ curves. 
From our simulation results, another broadening, around $m=130$, may be predicted.

\end{document}